\documentclass[prl,aps,twocolumn,preprintnumbers,showpacs,superscriptaddress]{revtex4-1}
\usepackage{}
\usepackage{amssymb}
\usepackage{amsfonts}
\usepackage{graphics,graphicx,epsfig,bm,amsthm,amsmath,amssymb}
\usepackage{bm}
\usepackage{bbm}

\usepackage{ulem}
\usepackage{float}
\usepackage[a4paper,colorlinks=true,
linkcolor=blue,citecolor=blue,
pdfauthor={ },
pdftitle={ },
pdfsubject={ },
pdfkeywords={ }]{hyperref}

\begin{document}

\title{Resolving Remoter Nuclear Spins in a Noisy Bath by Dynamical Decoupling Design}

\author{Wenchao Ma}
\affiliation{Hefei National Laboratory for Physical Sciences at the Microscale and Department of Modern Physics, University of Science and Technology of China, Hefei, Anhui 230026, People's Republic of China}

\author{Fazhan Shi}
\affiliation{Hefei National Laboratory for Physical Sciences at the Microscale and Department of Modern Physics, University of Science and Technology of China, Hefei, Anhui 230026, People's Republic of China}
\affiliation{Synergetic Innovation Center of Quantum Information and Quantum Physics, University of Science and Technology of China, Hefei, Anhui 230026, People's Republic of China}

\author{Kebiao Xu}
\affiliation{Hefei National Laboratory for Physical Sciences at the Microscale and Department of Modern Physics, University of Science and Technology of China, Hefei, Anhui 230026, People's Republic of China}

\author{Pengfei Wang}
\affiliation{Hefei National Laboratory for Physical Sciences at the Microscale and Department of Modern Physics, University of Science and Technology of China, Hefei, Anhui 230026, People's Republic of China}
\affiliation{Synergetic Innovation Center of Quantum Information and Quantum Physics, University of Science and Technology of China, Hefei, Anhui 230026, People's Republic of China}

\author{Xiangkun Xu}
\affiliation{Hefei National Laboratory for Physical Sciences at the Microscale and Department of Modern Physics, University of Science and Technology of China, Hefei, Anhui 230026, People's Republic of China}

\author{Xing Rong}
\affiliation{Hefei National Laboratory for Physical Sciences at the Microscale and Department of Modern Physics, University of Science and Technology of China, Hefei, Anhui 230026, People's Republic of China}
\affiliation{Synergetic Innovation Center of Quantum Information and Quantum Physics, University of Science and Technology of China, Hefei, Anhui 230026, People's Republic of China}

\author{Chenyong Ju}
\affiliation{Hefei National Laboratory for Physical Sciences at the Microscale and Department of Modern Physics, University of Science and Technology of China, Hefei, Anhui 230026, People's Republic of China}
\affiliation{Synergetic Innovation Center of Quantum Information and Quantum Physics, University of Science and Technology of China, Hefei, Anhui 230026, People's Republic of China}

\author{Chang-Kui Duan}
\affiliation{Hefei National Laboratory for Physical Sciences at the Microscale and Department of Modern Physics, University of Science and Technology of China, Hefei, Anhui 230026, People's Republic of China}
\affiliation{Synergetic Innovation Center of Quantum Information and Quantum Physics, University of Science and Technology of China, Hefei, Anhui 230026, People's Republic of China}

\author{Nan Zhao}
\affiliation{Beijing Computational Science Research Center, Beijing 100084, China}
\affiliation{Synergetic Innovation Center of Quantum Information and Quantum Physics, University of Science and Technology of China, Hefei, Anhui 230026, People's Republic of China}

\author{Jiangfeng Du}
\email{djf@ustc.edu.cn}
\affiliation{Hefei National Laboratory for Physical Sciences at the Microscale and Department of Modern Physics, University of Science and Technology of China, Hefei, Anhui 230026, People's Republic of China}
\affiliation{Synergetic Innovation Center of Quantum Information and Quantum Physics, University of Science and Technology of China, Hefei, Anhui 230026, People's Republic of China}


\begin{abstract}
We experimentally resolve several weakly coupled nuclear spins in diamond using a series of novelly designed dynamical decoupling controls.
Some nuclear spin signals, hidden by decoherence under ordinary dynamical decoupling controls, are shifted forward in time domain to the coherence time range and thus rescued from the fate of being submerged by the noisy spin bath.
In this way, more and remoter single nuclear spins are resolved. Additionally, the field of detection can be continuously tuned on sub-nanoscale.
This method extends the capacity of nanoscale magnetometry and may be applicable in other systems for high-resolution noise spectroscopy.
\end{abstract}
\maketitle

\section{I. Introduction}
Detection of single nuclear spins is an outstanding issue in magnetic resonance spectroscopy and imaging. It would be beneficial to molecular structure analysis, and will have a far-reaching impact on chemistry, biology, and medicine \cite{rugar04, Rugar2009PNAS}. Besides, nuclear spins are valuable as quantum registers for their long coherence times, particularly in relation to diamond defects \cite{dutt07, neumann08, jiang09, ladd10, maurer12, Wrachtrup2014Nature}. However, it is challenging to detect single nuclear spins owing to their weak magnetic moments and the noisy environment of solids.
Recently, single nuclear spin detection has been studied in various systems, including electrical transport measurements of $^{159}$Tb in a single-molecule magnet \cite{Vincent2012Nature} and all-electrical detection of $^{31}$P in silicon \cite{Morello2013Nature}. Additionally, using an individual negatively charged nitrogen-vacancy (NV) center in diamond, numerous protons in organic samples \cite{mamin13, Wrachtrup2013Science}, four $^{29}$Si in silica \cite{jelezko2014}, and a single surface proton \cite{lukin2014} have been detected.

In this work, the single electron spin of NV center in diamond is used as a magnetic sensor [Fig.~\ref{diamond}(a)] \cite{taylor08, maze08, balasubramanian08}.
Strongly coupled nuclear spins (with coupling strength exceeding the electron spin dephasing rate $\sim 1/T_2^*$) can be well resolved through the electron spin level splitting \cite{jelezko04, smeltzer11, jacques12}.
To resolve weakly coupled nuclear spins (with coupling strength comparable or less than the electron spin dephasing rate), the environmental noise has to be suppressed and the sensor sensitivity for nuclear spin signals has to be enhanced.
Dynamical decoupling (DD) methods \cite{viola99}, such as the widely used Carr-Purcell-Meiboom-Gill (CPMG) sequences and XY sequences, are the most commonly used tools at present. By periodically flipping central electron spins, weakly coupled single $^{13}$C nuclear spins \cite{zhaonano12, Lukin2012PRL, taminiau12} and clusters \cite{zhao11, Du2014NatPhys} around NV centers have been detected.
In this work, we apply novel DD controls \cite{Zhao2014arxiv} which have advantages in resolving remote single nuclear spins.
Applying the well-designed DD control sequences, some nuclear spin signals greatly diminished by decoherence are shifted forward in time domain to the coherence range, and thus rescued from the fate of being submerged by the noisy spin bath. With this benefit, two remoter nuclear spins, nearly silent under CPMG controls, pronounce their existence under appropriately designed DD controls in our experiment. Besides, the resolutions for characterizing some single nuclear spins can be improved using this method.

\section{II. Theory}
\begin{figure}
\centering
\includegraphics[width=1\columnwidth]{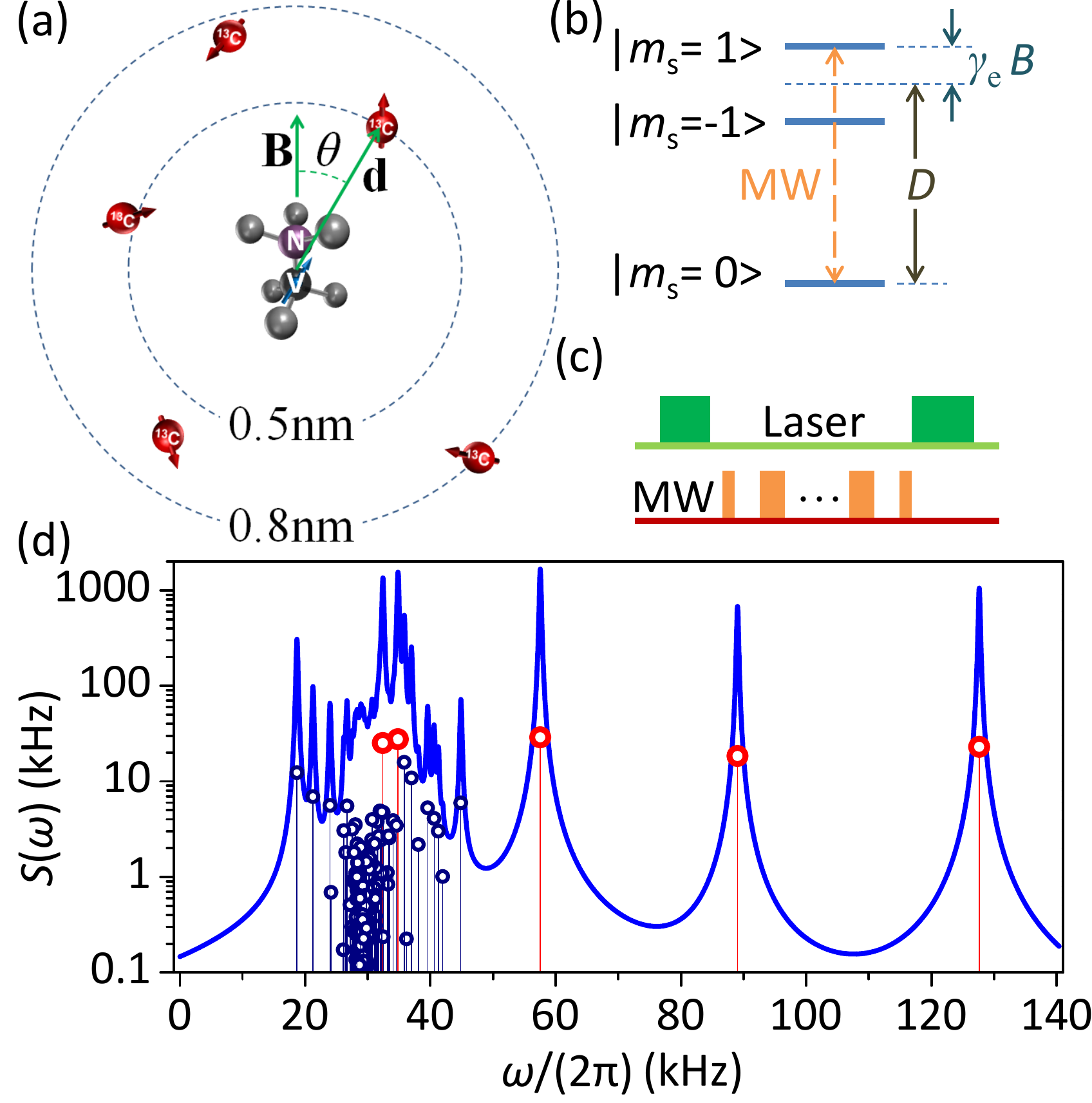}
    \caption{(color online). Sketch of the experimental system and methods.
   (a) The experimental system consists of an NV center and several weakly coupled $^{13}$C nuclear spins.
   (b) Ground state energy levels of an negatively charged NV center in an external static magnetic field \textbf{B}, which is 27 G and parallel to the NV axis. The zero field splitting $D$ is 2870 MHz. We encode the quantum transition $\vert m_s=0\rangle\leftrightarrow\vert m_s=1\rangle$ as a probe qubit and manipulate it by resonant microwave pulses.
   (c) The pulse sequences contain two green laser pulses for initialization and readout of the electron spin state with microwave pulses in between to control the spin. (d) Noise spectrum of the $^{13}$C nuclear spin bath. The five red circles correspond to the five single nuclear spins resolved in the experiments. The blue circles correspond to other bath spins, which are randomly generated by a simulating program considering the experimental conditions.
 }
    \label{diamond}
\end{figure}

The dynamics of nuclear spins around an NV center results in the modulation of the electron spin coherence and causes the decoherence effect \cite{Lukin2008PRB, Liu2012PRB}. This phenomenon can be viewed in a semiclassical picture, where the coherence $L(t)$ of a two-level system in a time-dependent noise field is approximately expressed as \cite{sarma08}
\begin{equation}\label{eq:coherence}
L(t) \sim \exp \left[ { - \int_0^\infty  {\frac{{d\omega }}{\pi }\frac{{S(\omega )}}{{{\omega ^2}}}} F(\omega t)} \right],
\end{equation}
with $t$ being the total evolution time, $S(\omega)$ the noise spectrum [Fig.~\ref{diamond}(d)], and $F(\omega t)$ the filter function [Figs.~\ref{dd-3}(b-d)] associated to the applied control pulse sequence. DD controls make the electron spin selectively sensitive to signals at specific frequencies and suppress unwanted background noise according to the features of $F(\omega t)$.
Nuclear spins around the NV center induce discrete peaks on $S(\omega)$ as Fig.~\ref{diamond}(d) shows \cite{supplement}, and thus can be experimentally identified via the characteristic dips on the electron spin coherence \cite{zhaonano12, Zhao2014arxiv}. Note that the semiclassical treatment here is qualitative and heuristic, and the quantitative and rigorous calculation should be based on quantum mechanics.

%
The filter function $F(\omega t)$ encapsulates the influence of DD control on decoherence, so it relates to the positions and magnitudes of nuclear spin signals in the time domain.
The filter function $F(\omega t)$ of a typical DD control has a series of peaks, which center at $\omega t/(2\pi)=(2k-1)n/2$ ($k = 1, 2, 3, \cdots $) for an $n$-pulse CPMG control (referred to as CPMG-$n$). As the total evolution time $t$ increases, the peaks sweep over the noise spectrum from high frequencies to low frequencies.
When a peak of $F(\omega t)$ coincides with a discrete peak in the noise spectrum, a coherence dip appears as qualitatively indicated by Eq.~(\ref{eq:coherence}).
The depths of dips depend on the noise features of target nuclear spins and the peak heights of $F(\omega t)$. For a sequence with $n$ flipping pulses, the peak height scales as $n^2$.
In the case of small $n$, the peak is so short that some dips will be quite shallow and can hardly be identified.
In the opposite situation (large $n$), the peak is unnecessarily tall. The dips may be oscillating and malformed, and the signals from individual nuclear spins will be blurred.
So the peak heights of the filter functions should be properly adjusted.
However, conventionally used periodic DD controls have limited freedom in tuning $F(\omega t)$. Increasing the pulse number is a natural but rough way to alter $F(\omega t)$. In fact, we can design DD control sequences by changing pulse distributions in a systematic way to carry out desirable tuning of $F(\omega t)$. By doing so, the signals of weakly coupled nuclear spins with different distances and orientations could be selectively optimized by such method.

A CPMG-$n$ sequence is $n$ repetitions of a ¡°$\frac{\tau}{2}$ - $\pi$ - $\frac{\tau}{2}$¡± unit.
In this paper, we expand the repetition unit to contain three $\pi$ pulses. The three $\pi$ pulses within a repetition unit is symmetric but non-uniformly distributed, i.e., the second $\pi$ pulse is located at the center of the unit, and the first and third $\pi$ pulses are separated from the second one by $3r\tau$ with the parameter $r$ ($0<r<0.5$) characterizing the relative positions of the pulses. In short, the new sequence is $n/3$ repetitions of a ¡°$\frac{3(1-r)\tau}{2}$ - $\pi$ - $3r\tau$ - $\pi$ - $3r\tau$ - $\pi$ - $\frac{3(1-r)\tau}{2}$¡± unit [Fig.~\ref{dd-3}(a)].
In this case, the filter function depends on both $n$ and $r$, and is expressed as
\begin{equation}\label{eq:filterfunction-3}
F_n^r(\omega t) = \frac{{8{{\sin }^2}\frac{{\omega t}}{2}}}{{{{\cos }^2}\frac{{3\omega t}}{{2n}}}}{\left( {{{\cos }^2}\frac{{3\omega t}}{{4n}} - \cos \frac{{3r\omega t}}{n}} \right)^2},
\end{equation}
where $n$ is divisible by $6$.
The dominant peaks of this function center at $\omega t/(2\pi)=(2k-1)n/6$ ($k = 1, 2, 3, \cdots $) for $r\neq 1/3$. In particular, the first one, which is especially responsible for nuclear spin detection, moves forward and comes out at $\omega t/(2\pi)=n/6$. The height of these peaks, corresponding to the amplification of the signals, depends on the value of $r$. The height of the $k$-th dominant peak [Fig.~\ref{dd-3}(f)] is
\begin{equation}\label{eq:peakheight}
h_n^k(r) = \frac{2}{9}{n^2}\left\{ {1 - 2\cos \left[ {(2k - 1)\pi r} \right]} \right\}^2.
\end{equation}
Such a degree of freedom provides advantages in nuclear spin detection. In the following we present the results for $n=30$.

\section{III. Experiments}
\begin{figure}
\centering
\includegraphics[width=1\columnwidth]{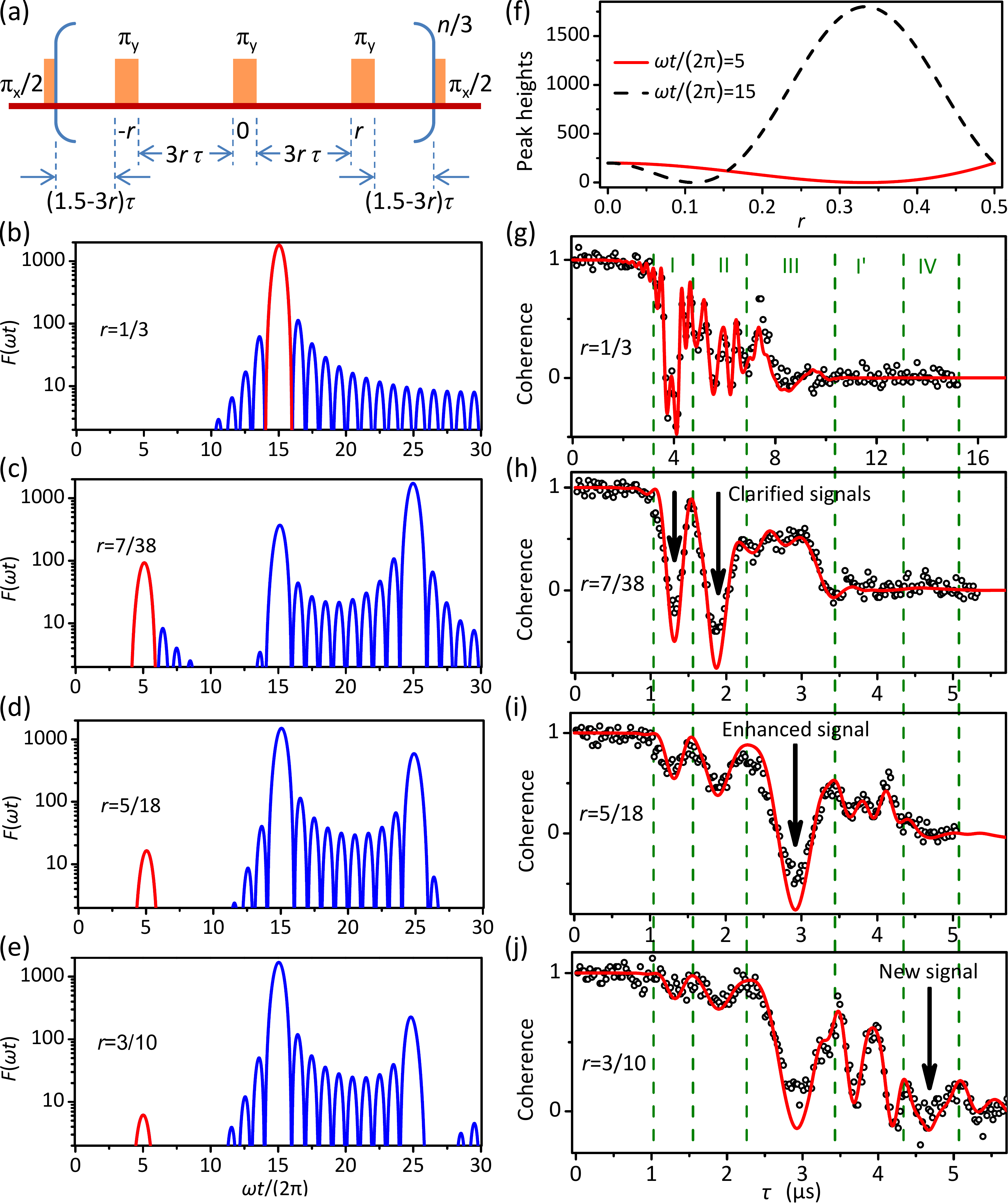}
\caption{(color online). Filter functions and coherence of the NV sensor for different controls.
 (a) Designed DD sequences. (b-e) Filter functions of DD controls with $r=1/3$, $7/38$, $5/18$, and $3/10$, respectively. The first dominant peaks are emphasized in red. (f) Heights of the dominant peaks as functions of $r$. The dashed and solid curves correspond to the peaks at $\omega t/(2\pi)=15$ and $5$, respectively. (g-j) Coherence under DD controls with $r=1/3$, $r=7/38$, $r=5/18$, and $3/10$, respectively. The horizontal coordinate $\tau$ donates $t/n$, where $t$ is the total free evolution time. Note that the range of the horizontal axes in (h-j) is $1/3$ of that in (g). The black circles represent experimental data. The red curves are calculations according to Eq.~S16 multiplied a decay factor.
 }\label{dd-3}
\end{figure}

We measured the spin coherence of an NV sensor embedded in a bulk diamond with nitrogen impurity of low concentration ($<5$ ppb) and $^{13}$C isotope of natural abundance $(1.1\%)$.
Several weakly coupled $^{13}\text{C}$ nuclear spins [Fig.~\ref{diamond}(a)] impose a.c. magnetic noise on the NV sensor with characteristic frequencies that depend on the external static magnetic field and hyperfine couplings.
Typically, the a.c. noise from different nuclear spins induce different modulations on coherence curves of the NV sensor.
The sensor coherence under CPMG-$30$ (i.e., $r=1/3$) control is illustrated in Fig.~\ref{dd-3}(g).
The dips in the zones labeled by \textit{I}, \textit{II}, and \textit{III} originate from three distinct nuclear spins, respectively. 
We notice that, with CPMG-$30$ (i.e., $r=1/3$) control, the oscillating patterns in \textit{I} and \textit{II} blur the signals from individual nuclear spins.
By applying the designed DD control with $r=7/38$, oscillations on the dip are significantly suppressed and the resolution is thus improved [Fig.~\ref{dd-3}(h)]. This allows for extracting the coupling information of single nuclear spins straightforwardly from each dips.

To selectively amplify the signal in \textit{III}, the designed DD control with $r = 5/18$ is applied. The dips in \textit{I}' also arise, and they originate from the same nuclear spin as that in \textit{I}. Under the designed DD control with $r=3/10$, a new coherence dip in \textit{IV} emerges [Fig.~\ref{dd-3}(j)]. This dip, not resolved under CPMG control, is due to two remoter nuclear spins \cite{supplement}. 
To sum up, the signals of nuclear spins for $r\neq 1/3$ [Figs.~\ref{dd-3}(h-j)] move forward to 1/3 positions in the time domain relative to the case of $r=1/3$ [Fig.~\ref{dd-3}(g)], while the coherence times are not shortened as much, especially for $r=5/18$ and $3/10$.
This relaxes the decoherence restriction and makes more and remoter nuclear spins detectable.
\begin{figure}
\centering
\includegraphics[width=1\columnwidth]{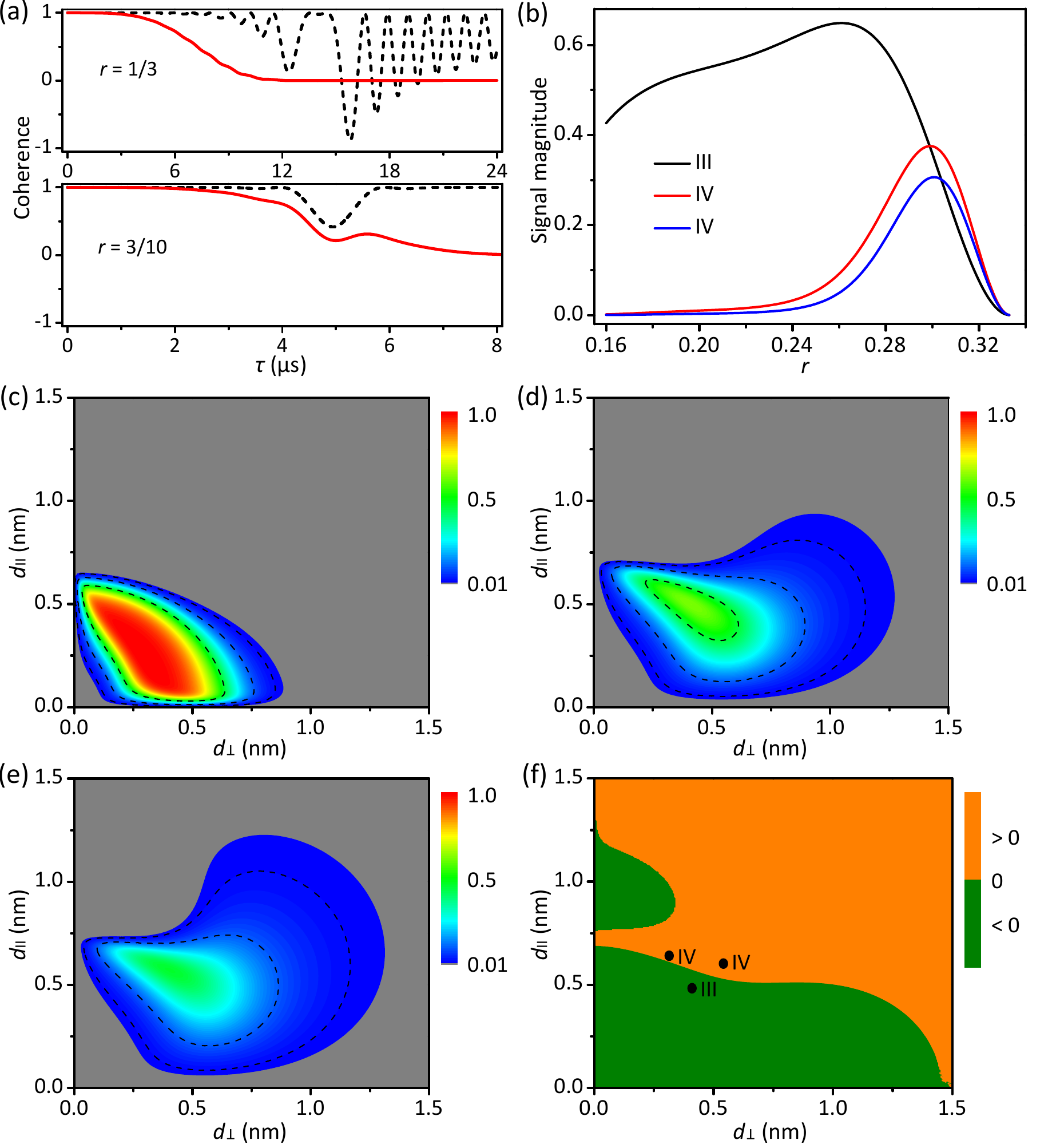}
\caption{(color online). Calculations.
(a) Coherence of the NV sensor with a nuclear spin nearby under DD controls with $r=1/3$ (upper) and $3/10$ (lower).
The decoherence is ignored in the dashed curves, while it is taken into account in the solid curves.
(b) Signal magnitudes of three nuclear spins resolved in experiment [Figs.~\ref{dd-3}(i,j)] as functions of $r$. (c-e) Signal magnitude maps of nuclear spins under DD controls with $r=1/3$, $5/18$, and $3/10$, respectively. The chroma represent the signal magnitudes, and the three color bars are the same. The dashed contours from the outside in correspond to signal magnitudes of $0.02$, $0.1$, and $0.5$. The grey areas have the signal magnitudes below $0.01$. (f) Signal magnitude difference between the cases of $r=5/18$ and $3/10$ by subtracting the former [Fig.~\ref{benefit}(d)] from the latter [Fig.~\ref{benefit}(e)]. The three dots represent three nuclear spins resolved in experiment [Figs.~\ref{dd-3}(i,j)].
}\label{benefit}
\end{figure}

To show the advantage of well-designed DD controls more clearly, we take a nuclear spin with $d=1$nm and $\theta=60^{\circ}$ for instance, where $d$ and $\theta$ represent the NV-$^{13}$C distance and the inclination angle of the NV-$^{13}$C vector with respect to the magnetic field \textbf{B} [Fig.~\ref{diamond}(a)].
The calculations show clearly that the nuclear spin can be resolved by the designed DD control with $r=3/10$, but cannot by the original CPMG control (i.e., $r=1/3$) under the same magnetic field [Fig.~\ref{benefit}(a)].
In the following we elaborate the analysis by more calculations based on Eq.~(\ref{eq:coherence}). The $r$ dependence of the depths of coherence dips (regarded as signal magnitudes) for the three detected nuclear spins [Figs.~\ref{dd-3}(i,j)] is plotted in Fig.~\ref{benefit}(b) \cite{supplement}.
It shows that optimal values of $r$ vary among nuclear spins.
Under the assumption that only dipole-dipole couplings are present between the sensor and nuclei, and that the dipole moment of the sensor is point-like, the signal magnitude of a nuclear spin in any spacial position can be determined. In this way, signal magnitude maps are obtained.
The maps plotted in Figs.~\ref{benefit}(c-e) correspond to $r=1/3, 5/18$, and $3/10$, respectively. The coordinates represent the longitudinal distance ${d_\parallel } = d\cos \theta$ and transverse distance ${d_ \bot } = d\sin \theta$ between the NV center and nuclei.
The three maps indicate that designed DD controls with $r=5/18$ and $3/10$ surpass the undesigned one with $r=1/3$ in detecting remoter nuclear spins, and this is consistent with the experimental results in Figs.~\ref{dd-3}(g,i,j).
The discrimination between Figs.~\ref{benefit}(d) and Figs.~\ref{benefit}(e) is shown in Fig.~\ref{benefit}(f).
The DD control with $r=3/10$ is superior in the positive areas.
This difference map implies that remoter nuclear spins can be observed with $r=3/10$ rather than $r=5/18$, and is consistent with the contrast between Figs.~\ref{dd-3}(i) and \ref{dd-3}(j). It is evident that nuclear spins in different regions are selectively resolved through the adjusting process. The distances between the relevant nuclear spins are of the magnitude of several angstroms as schematically illustrated in Fig.~\ref{diamond}(a) \cite{supplement},
so the field of detection can be tuned on sub-nanoscale. This behavior may enable tomography of the nuclear spin environment.

\begin{figure}
\centering
\includegraphics[width=0.7\columnwidth]{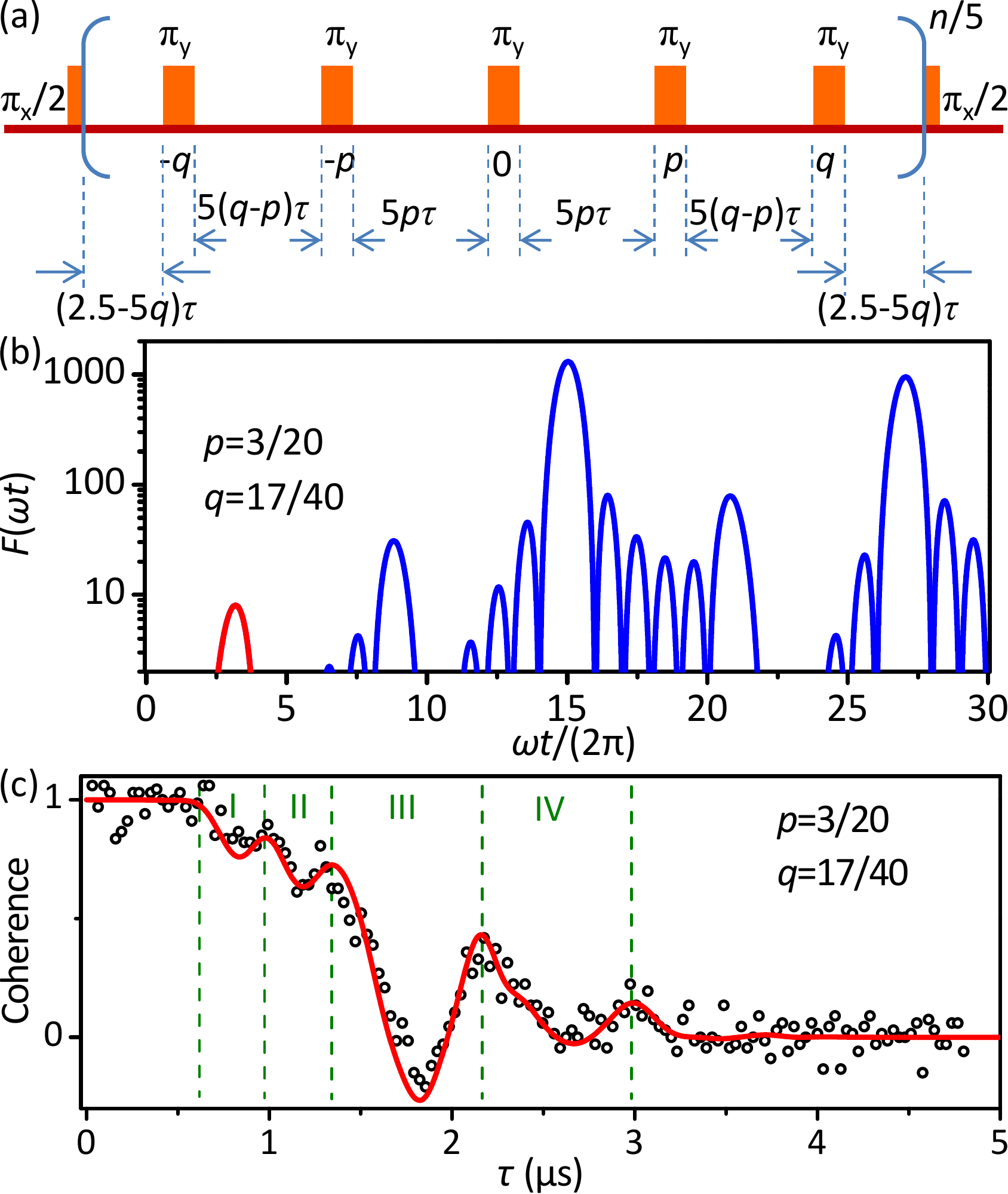}
\caption{(color online). Designed DD control with five pulses per unit.
 (a) Pulse sequence. (b) Filter functions of DD controls with $p=3/20$ and $q=17/40$. The first dominant peak is emphasized in red. (c) Coherence under DD controls with $p=3/20$ and $q=17/40$. The horizontal coordinate $\tau$ donates $t/n$, where $t$ is the total free evolution time. The black circles represent experimental data. The red curves are calculations according to Eq.~S16 multiplied a decay factor.
 }\label{dd-5}
\end{figure}

Furthermore, this method has variations. We can group every five pulses together as a unit and modify the sequences in a similar fashion [Fig.~\ref{dd-5}(a)]. The filter function [Fig.~\ref{dd-5}(b)] is
\begin{equation}\label{eq:filterfunction-5}
F_n^{p,q}(\omega t) = \frac{{8{{\sin }^2}\frac{{\omega t}}{2}}}{{{{\cos }^2}\frac{{5\omega t}}{{2n}}}}{\left( {{{\sin }^2}\frac{{5\omega t}}{{4n}} - \cos \frac{{5p\omega t}}{n} + \cos \frac{{5q\omega t}}{n}} \right)^2},
\end{equation}
where $n$ is divisible by 10 and $0<p<q<0.5$. The dominant peaks of this function center at $\omega t/(2\pi)=(2k-1)n/10$ ($k = 1, 2, 3, \cdots $) for $p\neq 1/5$ or $q\neq 2/5$.
The height of the $k$-th dominant peak is

\begin{equation}\label{eq:peakheight-5}
 h_n^k(p,q) =
 \frac{2}{{25}}{n^2}{\left\{ {1 - 2\cos \left[ {(2k - 1)\pi p} \right] + 2\cos \left[ {(2k - 1)\pi q} \right]} \right\}^2}.
\end{equation}
The signals of nuclear spins for $r\neq 1/3$ [Figs.~\ref{dd-3}(h-j)] move forward to 1/3 positions in the time domain relative to the case of $r=1/3$ [Fig.~\ref{dd-3}(g)]
The signals move forward to 1/5 positions of that in Fig.~\ref{dd-3}(g) in the time domain. Using well-designed five-piece DD controls, the unwanted signal in \textit{I}' of Fig.~\ref{dd-3}(j) can be eliminated, and the signals in \textit{III} and \textit{IV} stand out [Fig.~\ref{dd-5}(c)].

\section{IV. Conclusion}
In conclusion, we report the first experimental demonstration of DD design for identifying weakly coupled nuclear spins in a solid system. We show that properly designed DD controls outperform the conventionally used CPMG controls in resolving weakly coupled individual nuclear spins, resulting in more and remoter nuclear spins to be resolved.
This provides a route towards tomography of nuclear spin environment, and is inspiring in nanoscale magnetometry. It also extends the capability of using nuclear spins as quantum registers in quantum information processing. Additionally, the principle of DD design may be applicable in other systems, such as trapped ion and superconducting qubit, for high-resolution noise spectroscopy.

\section{ACKNOWLEDGMENTS}
This work was supported by the 973 Program (Grant Nos. 2013CB921800, 2014CB848700), the NNSFC (Grant Nos. 11227901, 91021005, 11275183, 11104262, 31470835, 11421063), the CAS (Grant No. XDB01030400), the Fundamental Research Funds for the Central Universities.

\end{document}